\documentclass[12pt,preprint]{aastex}

\shorttitle{Long-term variations and Rossby waves}
\shortauthors{Zaqarashvili et al.}

\begin{document}

\title{Long-term variation in the Sun's activity caused by magnetic Rossby waves in the tachocline}

\author{Teimuraz V. Zaqarashvili\altaffilmark{1,5},
Ramon Oliver\altaffilmark{2}, Arnold Hanslmeier\altaffilmark{3}, Marc Carbonell\altaffilmark{4}, Jose Luis Ballester\altaffilmark{2}, Tamar Gachechiladze\altaffilmark{5}, and Ilya G. Usoskin\altaffilmark{6}}

\altaffiltext{1}{Space Research Institute, Austrian Academy of Sciences,
Schmiedlstrasse 6, 8042 Graz, Austria. Email: teimuraz.zaqarashvili@oeaw.ac.at}
\altaffiltext{2}{Departament de F\'{\i}sica, Universitat de les Illes Balears, E-07122, Palma de Mallorca, Spain} \altaffiltext{3}{Institute f\"ur Physik, Geophysik Astrophysik und Meteorologie, University of Graz, Univ.-Platz 5, 8010 Graz, Austria}
\altaffiltext{4}{Departament de Matem\`{a}tiques i Inform\`{a}tica, Universitat de les Illes Balears, E-07122, Palma de Mallorca, Spain}
\altaffiltext{5}{Abastumani Astrophysical Observatory at Ilia State University, Tbilisi, Georgia}
\altaffiltext{6}{Sodankyl{\"a} Geophysical Observatory and ReSoLVE Centre of Excellence, University of Oulu, 90014 Finland}

\begin{abstract}
Long-term records of sunspot number and concentrations of cosmogenic radionuclides (10Be and 14C) on the Earth reveal the variation of the Sun's magnetic activity over hundreds and thousands of years. We identify several clear periods in sunspot, 10Be, and 14C data as 1000, 500, 350, 200 and 100 years. We found that the periods of the first five spherical harmonics of the slow magnetic Rossby mode in the presence of a steady toroidal magnetic field of 1200-1300 G in the lower tachocline are in perfect agreement with the time scales of observed variations. The steady toroidal magnetic field can be generated in the lower tachocline either due to the steady dynamo magnetic field for low magnetic diffusivity or due to the action of the latitudinal differential rotation on the weak poloidal primordial magnetic field, which penetrates from the radiative interior. The slow magnetic Rossby waves lead to variations of the steady toroidal magnetic field in the lower tachocline, which modulate the dynamo magnetic field and consequently the solar cycle strength. This result constitutes a key point for long-term prediction of the cycle strength. According to our model, the next deep minimum in solar activity is expected during the first half of this century.

\end{abstract}


\section{Introduction}\label{intro}

Solar activity, which is manifested through huge energy releases in terms of coronal mass ejections and solar flares, basically determines the plasma conditions of near-Earth and interplanetary space, and possibly influences Earth's climate \citep{Haigh2007}. Therefore, the prediction of solar activity is of paramount importance for our future scientific-technical strategy. The activity undergoes 11-year oscillations known as Schwabe cycles \citep{Schwabe1844} generally interpreted in terms of dynamo models \citep{Charbonneau2005}. However, long-term records of sunspot number \citep{Gleissberg1939,Hill1977} and concentrations of cosmogenic radionuclides on the Earth \citep{Stuiver1989,Suess1980,Solanki2004,Usoskin2004,Steinhilber2012} reveal the variation of the Sun's magnetic activity over hundreds and thousands of years.

The observed long-term variation in solar activity is still a mystery \citep{Hathaway2010}. Three different explanations for the variation have been proposed. The first one suggests that the nonlinear solar dynamo permits a transition from periodic oscillations to chaotic behavior with a long period modulation of magnetic field strength \citep{Weiss1984}. The second mechanism considers random fluctuations in the dynamo parameters, which may mimic the main features of the long-term variations in the activity \citep{Choudhuri2012}. The third proposal is that the tidal influence of solar system planets may modulate the temperature gradient and consequently the magnetic storage capacity in the solar tachocline \citep{abreu2012}. The first two mechanisms may account for the occurrence of Grand Minima, but fail to reproduce the observed long-term modulation of solar cycles. The third one, the planetary hypothesis (a long standing problem), has been intensively discussed in the last years by the scientific community \citep{Charbonneau2013}, but has been criticized due to the inappropriate statistical analysis \citep{Cameron2013,Poluianov2014} or data artefact \citep{Cauquoin2014}.

The long-term variation of solar activity is probably connected to the tachocline, a thin layer below the solar convection zone, which is a very important place for the solar dynamo and angular momentum redistribution \citep{Spiegel1992}. The tachocline is believed to consist of a lower strongly stratified and an upper overshoot weakly stratified layers \citep{Gilman2000}. The convective motions from the convection zone penetrate the upper overshoot layer, while the lower layer is much more stable and the ordinary magnetohydrodynamic (MHD) shallow water equations of \citet{Gilman2000} can be safely applied there.

The conservation of total vorticity in hydrodynamic shallow water equations leads to the appearance of Rossby or planetary waves in the rotating frame, which govern the large-scale dynamics of the Earth's atmosphere (and oceans) causing the formation of cyclones/anticyclones and hurricanes. Inclusion of horizontal magnetic field splits the ordinary hydrodynamic Rossby waves into fast and slow magnetic Rossby waves in a magnetized plasma  \citep{zaqarashvili2007,zaqarashvili2009}. Fast magnetic Rossby waves are responsible for the observed Rieger type-periodicity (155-160 days) in solar activity after the interaction with the dynamo magnetic field in the tachocline \citep{zaqarashvili2010}. However, slow magnetic Rossby waves could have very long time scales compared to the solar rotation and solar cycle periodicity. In this letter, we consider the slow magnetic Rossby modes in the lower strongly stable layer of the solar tachocline and show their connection with long-term variations in solar activity.

\section{Slow magnetic Rossby waves and long periods in solar activity}

We consider the spherical coordinate system $(r, \theta,\phi)$, where $r$ is the radial coordinate, $\theta$ is the co-latitude and $\phi$ is the longitude. The unperturbed magnetic field is assumed to be toroidal i.e. ${\bf B}=B_{\phi}(\theta) {\bf \hat e_{\phi} }$. Then, the consideration of rigid rotation and a homogeneous magnetic field, $B_{\phi}=B_0=const$, leads to the following dispersion relation for magnetic Rossby waves in the rotating frame \citep{zaqarashvili2007}
\begin{equation}
\label{om} \omega^2+{{2\Omega_0m}\over {n(n+1)}}\omega+ {{B^2_0m^2}\over {4 \pi \rho R^2}}{{2-n(n+1)}\over {n(n+1)}}=0,
\end{equation}
where $\omega$ is the frequency of the spherical harmonic, $\Omega_0$  is the equatorial angular velocity, $R$ is the distance from the solar centre, $\rho$ is the medium density, $m$ and $n$ are toroidal and poloidal wavenumbers, respectively. If ${{B^2_0}/({4 \pi \rho \Omega^2_0R^2}})\ll1$, which is easily satisfied in the solar tachocline, then the low order spherical harmonics of slow magnetic Rossby modes are described by the dispersion relation
\begin{equation}
\label{sph} \omega_{mn}=-m\Omega_0{{B^2_0}\over {4 \pi\rho\Omega^2_0R^2}}{{2-n(n+1)}\over {2}},
\end{equation}
where $\omega_{mn}$ is the frequency of the spherical harmonic. Each spherical harmonic with different $m$ and $n$ has a different frequency (and hence period) depending on the plasma parameters; mostly on the magnetic field strength, $B_0$.

In order to find the correlation between slow magnetic Rossby waves and the long-term variation in solar activity, the periods of spherical harmonics with different magnetic field strength should be compared to the time scales found in observational data. We performed a spectral analysis of yearly mean total sunspot number from 1700 to 2013, and found strong peaks at 100 and 180 years in the oscillation spectrum (Figure 1, upper panel). The periods can be identified with the well known Gleissberg \citep{Gleissberg1939,Hathaway2010} and Suess \citep{Suess1980} cycles, respectively. Longer oscillatory periods cannot be safely defined due to the short length of the data series ($\sim$ 310 years). To find longer periods, we used solar activity reconstructed from cosmogenic isotopes \citep{Vonmoos2006,Usoskin2007} 10Be and 14C measured in the Greenland GRIP ice core and in tree rings, respectively, during the past 10 millennia. Both data sets show different periodicities \citep{Usoskin2004,Steinhilber2012,Hanslmeier2013}. In order to select reliable periods, we put forward a novel approach and overlapped the spectra of both data sets (Figure 1, lower panel). Then we selected four well defined periods of 1000, 500, 350 and 200 years in both data sets. 10Be data also shows power at the periods of 700-800 years, but there are no corresponding periods in 14C data. Combining the power spectra of sunspot numbers and cosmogenic isotopes, we define the reliable periods in solar activity as $\sim$ 1000, $\sim$ 500, $\sim$ 350, $\sim$ 200 and $\sim$ 100 years.

Using the plasma parameters at the tachocline ($\Omega_0$ =2.7$\cdot$10$^{-6}$ s$^{-1}$, $\rho$ =0.2 g$\cdot$cm$^{-3}$, $R$ = 5$\cdot$10$^{10}$ cm) and the magnetic field strength of 1200 G, we find that the periods of the modes $m$=1, $n$ =2, 3, 4, 5, 6 as calculated from Eq. (2) are $\sim$ 1170, $\sim$ 470, $\sim$ 260, $\sim$ 170 and $\sim$ 110 years, respectively, which show perfect agreement between theoretical and observed periods (Figure 2). Therefore, we suggest that the observed periods in solar activity are the consequence of the first five spherical harmonics of slow magnetic Rossby modes in the lower tachocline.

The magnetic field strength of 1200 G, considered above, is much smaller than the estimated value of dynamo generated magnetic field in the tachocline (10$^{4}$-10$^{5}$ G). Nevertheless, the toroidal component of the reversing dynamo field is zero when averaged over a time interval longer than the solar cycle length, because of periodic sign reversals in consecutive cycles. Therefore, it probably cannot influence the long-period dynamics of slow magnetic Rossby waves. However, there are two different mechanisms, which may be responsible for the appearance of weak steady (nonreversing) toroidal magnetic field in the lower tachocline or upper radiative interior. First, \citet{Dikpati2006} showed that a steady (nonreversing) dynamo for a low enough magnetic diffusivity $\leq$ 10$^7$ cm$^2$ s$^{-1}$ may generate steady toroidal magnetic field with the strength of $\sim$ 1 kG in the lower tachocline. This is consistent to our requirements. Second, the primordial magnetic field, which remained in the radiative interior after the Sun's collapse, may penetrate into the tachocline and affect Rossby waves. Both, nonreversing dynamo and the primordial fields may be responsible for the observed Gnevyshev-Ohl 22-year rule \citep{Gnevyshev1948}, which shows that odd sunspot cycles are generally stronger than even cycles, due to the break of polarity symmetry of the dynamo state \citep{Boyer1985}: the reversing dynamo magnetic field is aligned and anti aligned with respect to the toroidal component of steady field in consecutive cycles leading to the consequence of strong and weak cycles, respectively. Similarly, the periodical variation caused by slow magnetic Rossby waves in the steady magnetic field can influence the strength of the dynamo magnetic field and consequently the cycle strength.

The primordial magnetic field strength was estimated as 1000 G deep in the radiative interior and 1 G just beneath the tachocline \citep{Gough1998}. This is the poloidal magnetic field, which probably has negligible influence on the cycle strength. Helioseismology shows that the radiative interior rotates almost uniformly, therefore the magnetic field will remain poloidal in most of its part. On the other hand, the convection zone and the tachocline have differential rotation expressed as
\begin{equation}
\label{omega} \Omega=\Omega_0 (1-s_2 \cos^2\theta-s_4\cos^4\theta),
\end{equation}
where $s_2$ and $s_4$ are parameters determined by observations whose values at the solar surface (and in the middle tachocline) are about 0.14. \citet{Schou1998} constructed (see Figure 7 in that paper) the radial dependence of solar rotation at different latitudes along the whole convection zone and radiative interior as inferred by helioseismology. The authors determined the center of the tachocline at 0.7-0.71 $R_\odot$ with the negligible thickness of 0.05 $R_\odot$, therefore the lower boundary of the tachocline is located somewhere at 0.68 $R_\odot$. A careful look at Figure 7 of \citet{Schou1998} shows that $\Omega$ is 445 nHz at $\theta$=0 and 415 nHz at $\theta$=60 degree near the tachocline center (0.7 $R_\odot$), which gives the differential rotation parameter $s_2$=0.1. This is close to the main value of differential rotation in the tachocline. On the other hand, $\Omega$ is 440 nHz at $\theta$=0 and 435 nHz at $\theta$=60 degree near the lower boundary of the tachocline (at 0.68 $R_\odot$), which gives $s_2$=0.015. Therefore, a small differential rotation is still present in the lower tachocline. This differential rotation may eventually lead to the generation of a toroidal component from the poloidal primordial field, which penetrates from the upper part of the radiative envelope into the lower part of the tachocline. The induction equation allows to estimate  the generated toroidal field as $B_{\phi}/t_0\approx B_{\theta}(\partial \Omega/\partial \theta)$, where $t_0$ is the characteristic time (taken as the Alfv\'en time $R\sqrt{4\pi\rho}/B_{\theta}$) and $B_{\theta}$ is the poloidal component of the primordial magnetic field. Then the toroidal magnetic field is
\begin{equation}
\label{magnetic} B_{\phi}\sim {\hat B_0} \cos \theta \sin \theta,
\end{equation}
where ${\hat B_0}$ is 4300 G for $s_2$=0.01 and 1500 G for $s_2$=0.0035. The strength of the estimated magnetic field coincides with the value required for slow magnetic Rossby waves to explain the observed periodicity. It should be noted that the equipartition value of steady toroidal magnetic field with the kinetic energy of turbulent pumping from the convection zone expressed as $B_{eq}\approx v \sqrt{4\pi\rho}$, where $v$ is the turbulent velocity of convective cells (0.01 km$\cdot$s$^{-1}$ near the base of the convection zone according to the standard mixing length theory of convection), is also estimated as 1500 G.

Therefore, the strength of the steady toroidal magnetic field in the lower tachocline generated either by steady dynamo for low magnetic diffusivity or by differential rotation acting on the primordial poloidal component can be safely taken as 1000-1500 G.

The solar cycle strength is defined by magnetic energy of dynamo layer, therefore it is connected to the square of magnetic field strength. Longitudinally averaged magnetic energy of toroidal magnetic field is proportional to $\langle (B_d+B_s)^2 \rangle_{\phi}=\langle B^2_d \rangle_{\phi}+\langle 2B_dB_s \rangle_{\phi}+\langle B^2_s \rangle_{\phi}$, where brackets denote averaging over $\phi$ from $0$ to $2\pi$, $B_d$ is the dynamo generated reversing magnetic field and $B_s$ is the steady nonreversing (primordial or steady dynamo) magnetic field. Spherical harmonics of slow magnetic Rossby waves lead to the toroidal dependence of steady magnetic field corresponding to the wave number, $m$. For example, $m=1$ mode leads to the dependence of $\cos\phi$ in the steady field. Then the longitudinally averaged magnetic energy is proportional to $B^2_d+B^2_{s0}/2$, where $B_{s0}$ is the amplitude of steady field. $B_{s0}$ has temporal variations due to the slow magnetic Rossby waves as discussed above, therefore it can be responsible for the long period modulation of longitudinally averaged magnetic energy and hence the solar cycle strength.

The latitudinal differential rotation in the lower part of the tachocline (or in the upper part of the radiative interior), even if small, may lead to magnetohydrodynamic instabilities in joint action with the toroidal magnetic field \citep{Gilman1997,Cally2003}. In order to find unstable harmonics, we use the general technique of Legendre polynomial expansion \citep{Longuet-Higgins1968}. 2D shallow water MHD equations in the presence of differential rotation (Eq. \ref{omega}) and a toroidal magnetic field (Eq. \ref{magnetic}) can be written in the rotating frame as \citep{zaqarashvili2010}

\begin{equation}\label{momentum-spectrum}
(\Omega_d - {\hat \omega})L\Psi + (2 - {d^2\over {d
\mu^2}}[\Omega_d(1-\mu^2)])\Psi - \mu \beta^2 L \Phi - 6\mu \beta^2\Phi =0,
\end{equation}
\begin{equation}\label{induction-spectrum}
(\Omega_d- {\hat \omega})\Phi=\mu \Psi,
\end{equation}
where
$$
L={{\partial }\over {\partial \mu}}(1-\mu^2){{\partial }\over
{\partial \mu}} - {m^2\over {1-\mu^2}}
$$
is the Legendre operator, $\Psi$ and $\Phi$ are the stream functions for velocity and magnetic field perturbations, $\mu=\cos \theta$ and
$$\Omega_d(\mu)=-s_2\mu^2-s_4\mu^4, \,\, {\hat \omega}={\omega\over \Omega_0},\,\,\, \beta^2={{B^2_0}\over
{{4\pi \rho \Omega^2_0R^2}}}.
$$
We expand $\Psi$ and $\Phi$ in infinite series of associated Legendre polynomials
\begin{equation}\label{legandre}
\Psi=\sum^{\infty}_{n=m}a_nP^m_n(\mu),\,\,\,\Phi=\sum^{\infty}_{n=m}b_nP^m_n(\mu),
\end{equation}
substitute them into Eqs.
(\ref{momentum-spectrum})-(\ref{induction-spectrum}) and, using a recurrence relation of Legendre polynomials, obtain algebraic
equations as infinite series \citep{zaqarashvili2010}. The dispersion relation for the infinite number of harmonics can be obtained when the infinite determinant of the
system is set to zero. In order to solve the determinant, we truncate the series at $n=75$ and solve the resulting polynomial in
$\omega$ numerically. The frequencies of different harmonics can be real or complex giving the stable or unstable character of a particular harmonic. Then differential rotation parameters $s_2=s_4=0.01$ (as estimated in the previous paragraph) and the magnetic field strength $B_0$ = 1200-1300 G lead to periods of unstable harmonics - which are close to the observationally obtained and analytically predicted values (Figure 2).

\section{Discussion and Conclusions}

It is shown in this letter that the spherical harmonics of slow magnetic Rossby waves have very long periods in the presence of a steady toroidal magnetic field near the lower boundary of the tachocline. The steady toroidal field can be generated in the lower tachocline either due to a steady dynamo in the case of low magnetic diffusivity \citep{Dikpati2006} or due to action of latitudinal differential rotation on poloidal primordial magnetic field. The strength of steady dynamo magnetic field in the lower tachocline is $\sim$ 1 kG in the case of low magnetic diffusivity ($\leq$ 10$^7$ cm$^2$ s$^{-1}$) \citep{Dikpati2006}. The poloidal primordial magnetic field \textbf{may} penetrate the lower part of the tachocline from the radiative interior and generate the toroidal component due to the small latitudinal differential rotation, which is present near the interface of the radiative interior and the tachocline \citep{Schou1998}. The differential rotation rate of the order of $s_2=0.003-0.01$ leads to the equipartition value of toroidal magnetic field of 1500-4000 G. In both cases of steady dynamo and primordial magnetic fields, the periods of spherical harmonics of slow magnetic Rossby waves correspond to the observed long periods in sunspot number and radionuclide data. We believe that this coincidence has reliable physical ground: slow magnetic Rossby modes cause periodical variations of steady nonreversing toroidal magnetic field in the lower tachocline, which in turn modify the dynamo magnetic field and hence the solar cycle strength. The lower part of the tachocline is stable contrary to the upper overshoot layer as the convection does not penetrate there. Therefore, the steady toroidal magnetic field and large-scale slow magnetic Rossby waves may survive over long time without significant influence.

 MHD instabilities in the solar tachocline cause the appearance of active longitudes depending on the toroidal wave number, $m$ \citep{Dikpati2005}. Then the magnetic Rossby waves may lead to the migration of active longitudes in the toroidal direction \citep{Berdyugina2003,Gyenge2014}. The phase speed of slow magnetic Rossby waves along the toridal direction is \begin{equation}
\label{sph} v_{ph}=-{{B^2_0}\over {4 \pi\rho\Omega_0R}}{{2-n(n+1)}\over {2}},
\end{equation}
which for $n=2$ and $B_0$=1200 G has the value of 8 cm s$^{-1}$. The speed is much slower compared to the surface speed of solar rotation, therefore it is hard to catch the migration by observations.

Slow magnetic Rossby waves, which reproduce the observed long periods in solar activity, can be used for the long-term forecasting of the activity level. We fit the sum of five sinusoidal functions (with periods of 1000, 500, 350, 200 and 100 years), which are actually the spherical harmonics of slow magnetic Rossby modes, to the data of yearly mean total sunspot number from 1700 to 2013 (Figure 3). The resulting curve fits rather well the long-term behavior of the activity. Our model predicts the next deep minimum of solar activity around year 2030, which means that the next two solar cycles will be rather weak, but then the activity will start to increase again.

We also fit the sum of four sinusoidal functions (1000, 500, 350, 200 years) to the 100-year averaged sunspot number for the last three millennia, reconstructed from 14C data (Figure 4). The fit is generally good except near the Sp\"orer minimum, where a remarkable anti phase relation is seen. This disagreement is probably caused due to two distinct modes in solar activity as recently suggested by \citet{Usoskin2014}.

The theory we have described provides a simple view of the coupling between the large-scale dynamics of the radiative interior and the tachocline. Dynamic interaction between rotation and steady (steady dynamo or primordial) toroidal magnetic field in the upper part of the radiative interior and the tachocline inevitably results in long periodicities due to slow magnetic Rossby modes, which accordingly modify the reversing dynamo magnetic field causing the observed long-term variations in solar activity. The scenario is a starting point for future sophisticated models with comprehensive study (including intense numerical simulations), which will enhance the accuracy of long-term solar activity forecast. The theory may also provide a hint for observed long-period modulations of magnetic activity in solar-like stars.

{\bf Acknowledgements} The work of TZ was supported by the Austrian "Fonds zur F\"{o}rderung der wissenschaftlichen Forschung" under project P26181-N27, by FP7-PEOPLE-2010-IRSES-269299 project- SOLSPANET and by Shota Rustaveli Foundation grant DI/14/6-310/12. I. G.U. acknowledges the ReSOLVE Centre of Excellence (Academy of Finland, project no 272157). Source of yearly mean sunspot numbers: WDC-SILSO, Royal Observatory of Belgium, Brussels.

\clearpage

\begin{figure}
\includegraphics[angle=-90,scale=.50]{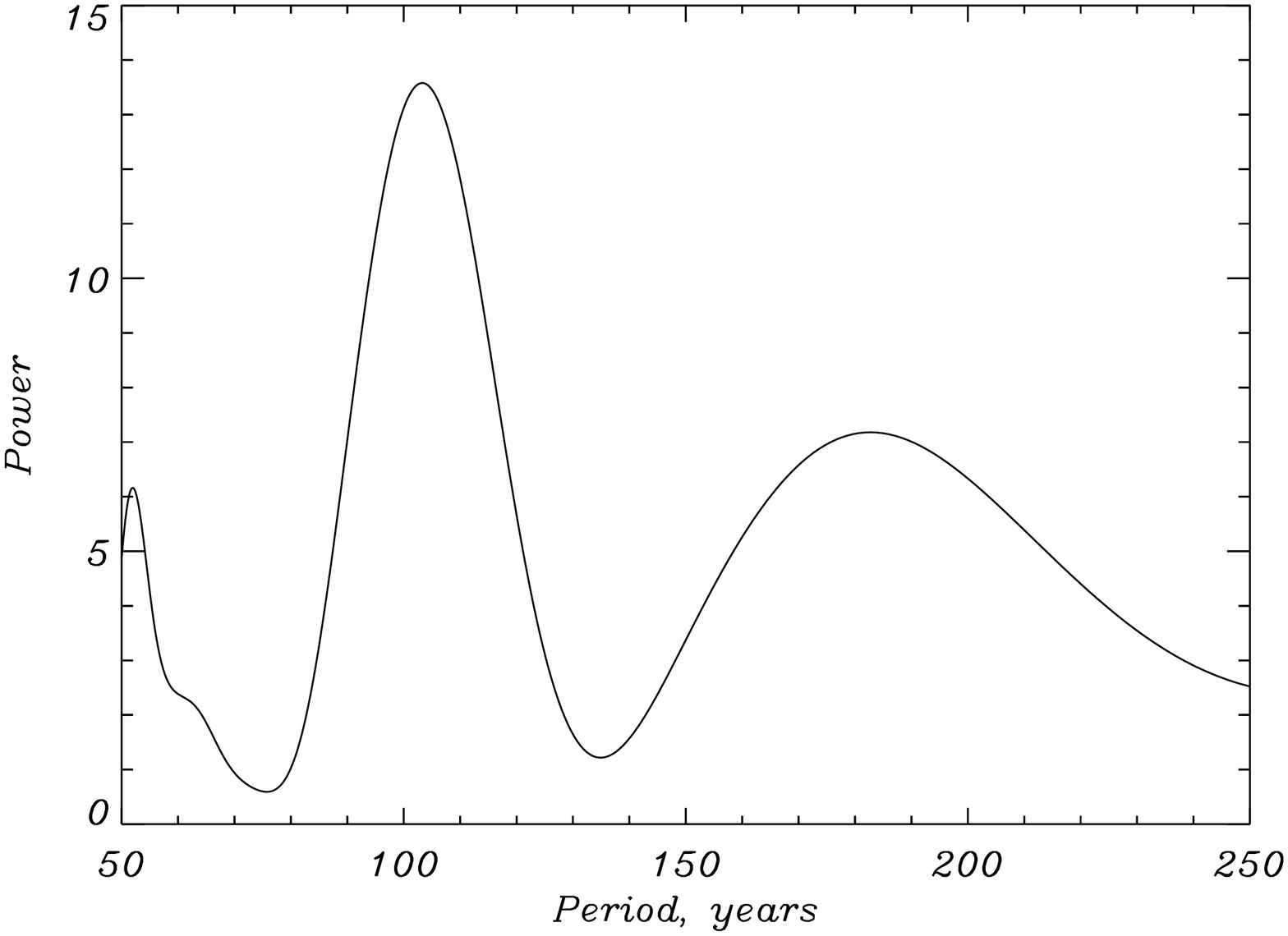}
\includegraphics[angle=-90,scale=.50]{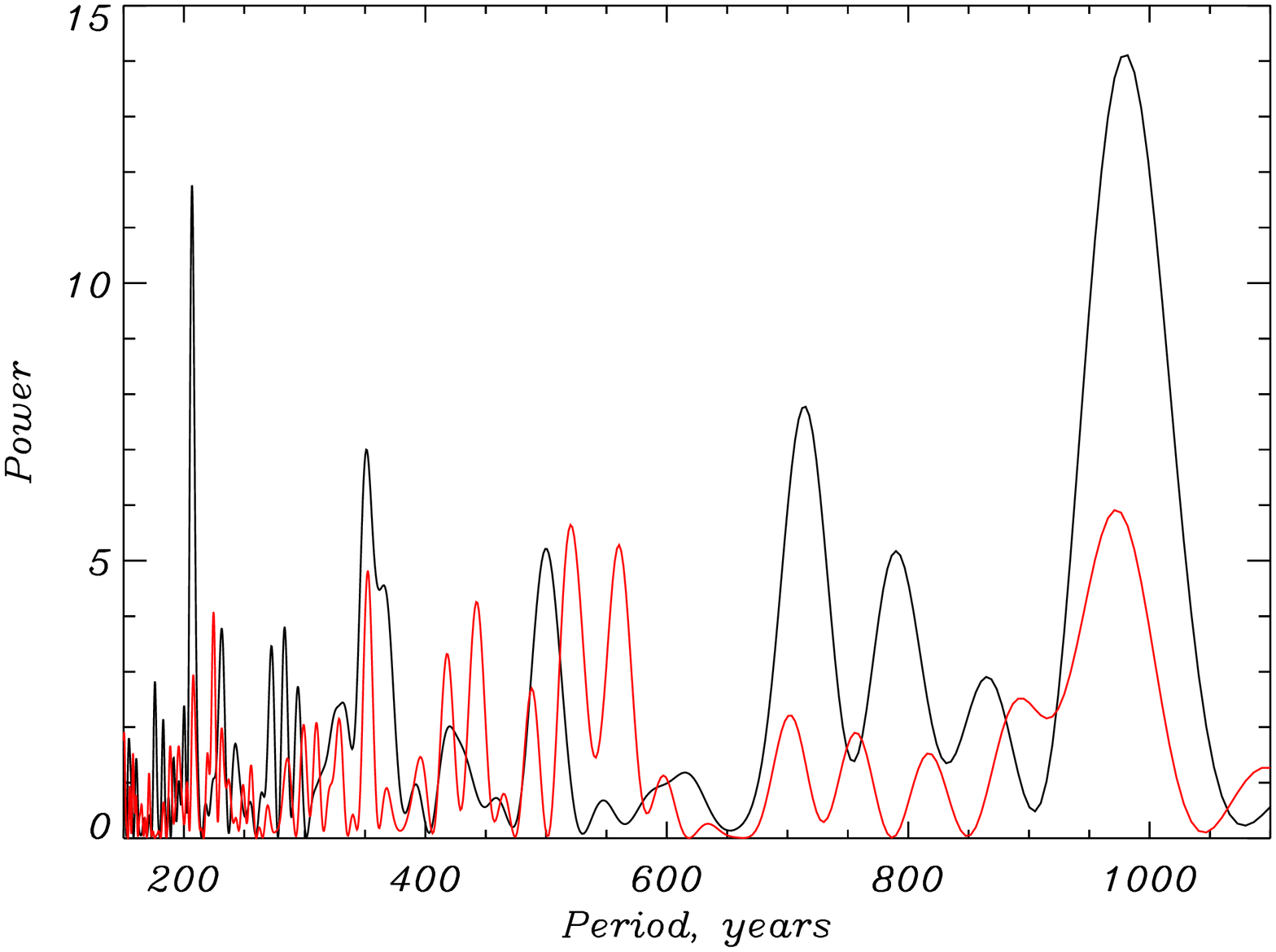}
\caption{Upper panel: Periodogram of yearly mean sunspot numbers. Large power values are concentrated around 180 and 100 years. Power is given in arbitrary units. Yearly mean sunspot data is downloaded from the World Data Center-SILSO, Royal Observatory of Belgium, Brussels. Lower panel: periodograms of reconstructed solar activity based on cosmogenic isotopes 10Be \citep{Vonmoos2006} (black line) and 14C \citep{Usoskin2007} (red line). Large power values in both data are concentrated around 1000, ~500, 350 and 200 years. Power is given in arbitrary units.}
\end{figure}

%

\clearpage

\begin{figure}
\epsscale{1}
\plotone{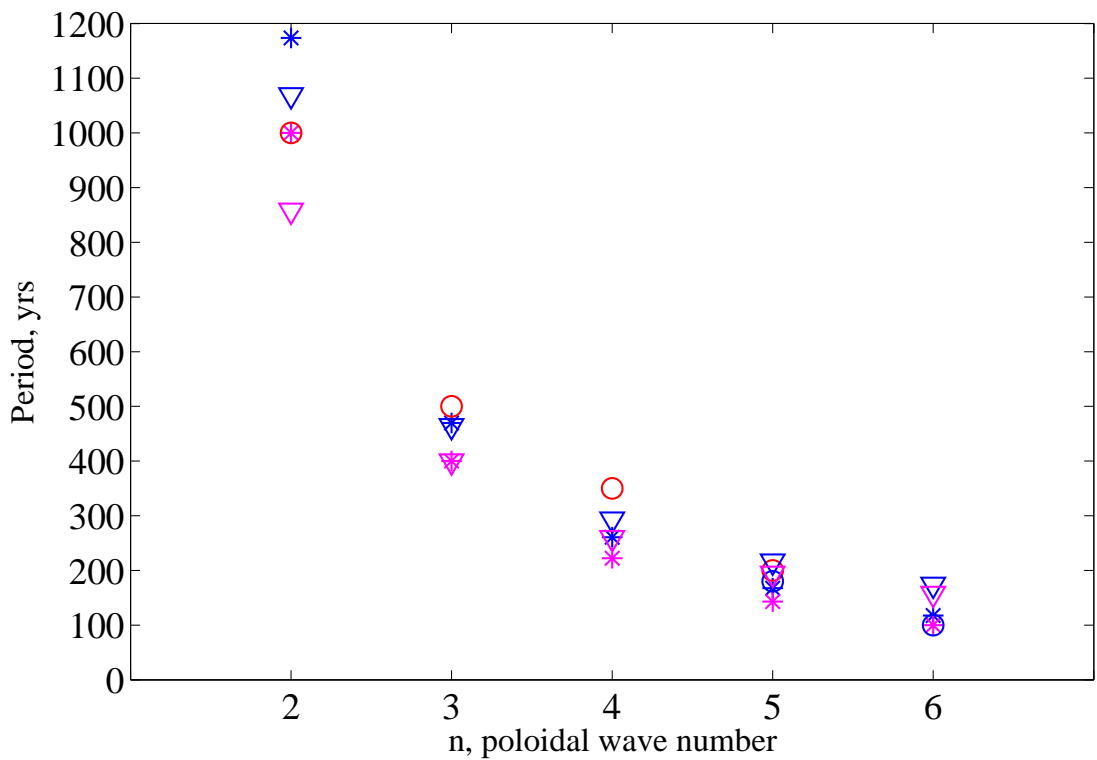}
\caption{Periods of solar activity and slow magnetic Rossby modes with different poloidal wave number, n. Blue and Red circles indicate the periods found by periodogram analysis of sunspot (Figure 1, upper panel) and radionuclide (Figure 1, lower panel) data, respectively. Blue and magenta asterisks indicate the periods of slow magnetic Rossby wave harmonics calculated from Eq. (2) for m=1 and n=2, 3, 4, 5, 6 for magnetic field strength of 1200 and 1300 G, respectively. Blue and magenta triangles indicate the periods of the first five unstable harmonics calculated by the Legendre polynomial expansion for magnetic field strength of 1200 and 1300 G, respectively. Here m=1 and the differential rotation parameters are $s_2=s_4=0.01$.  }\label{fig2}
\end{figure}

\clearpage

\begin{figure}
\includegraphics[angle=-90,scale=.70]{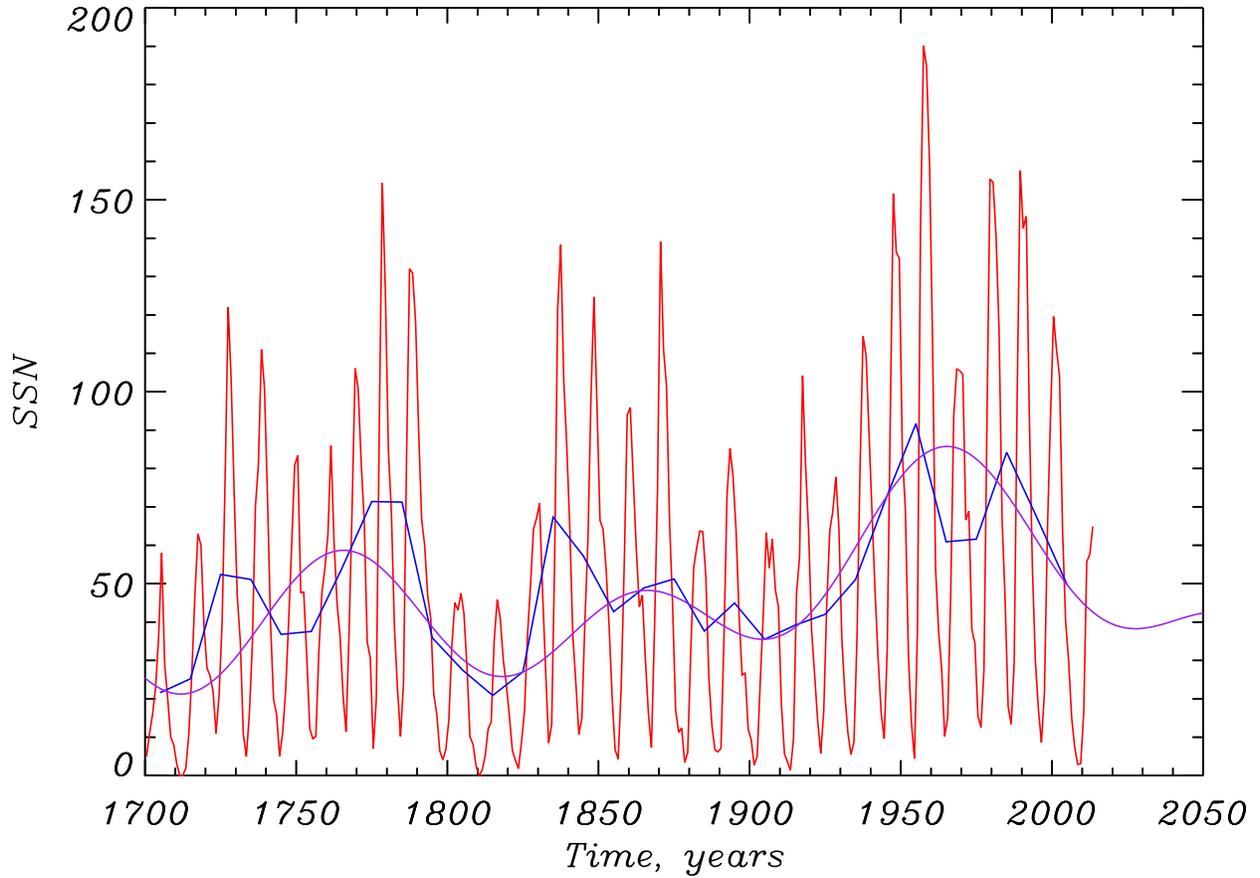}
\caption{Forecast of future solar cycle strength using magnetic Rossby wave theory. Red and blue lines show yearly and 10-year averaged sunspot numbers, respectively. Magenta line shows the fit of the sum of 5 sinusoidal functions with periods of 1000, 500, 350, 200 and 100 years.}
\end{figure}

%

\clearpage

\begin{figure}
\includegraphics[angle=-90,scale=.70]{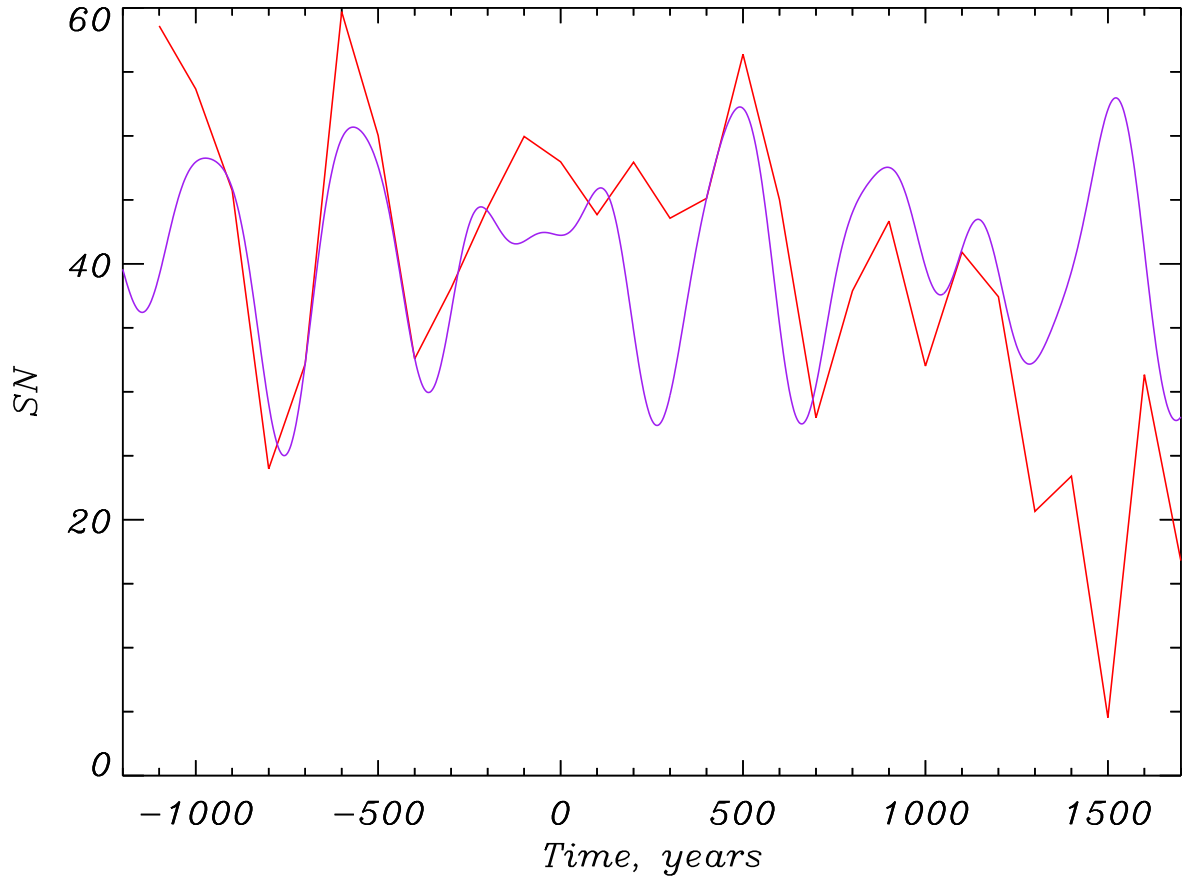}
\caption{Fit of magnetic Rossby wave harmonics to reconstructed sunspot number for the last three millennia. Red line shows 100-year averaged sunspot number reconstructed from 14C data \citep{Usoskin2014}. Purple line shows the fit of the sum of 4 sinusoidal functions with periods of 1000, 500, 350 and 200 years, which are the first four harmonics  of slow magnetic Rossby waves.}
\end{figure}


\end{document}